\documentclass[%
superscriptaddress,
 aps,prx,
 twocolumn,
 tightenlines,floatfix,nofootinbib]{revtex4-1}

\usepackage{graphicx}
\usepackage{dcolumn}
\usepackage{bm}
\usepackage{amsmath,amssymb}
\usepackage{natbib}
\usepackage{color,units}
\usepackage{float}
\usepackage{subfigure}
\usepackage[caption=false]{subfig}
\usepackage{aas_macros}

\def\Mtov{M_\mathrm{TOV}}
\def\tcol{t_{\mathrm {col}}}
\def\nn{\langle n \rangle}
\newcommand{\bilby}{{\sc Bilby}}
\newcommand{\dynesty}{{\sc dynesty}}
\newcommand{\gwpop}{{\sc gwpopulation}}
\def\swift{\textit{Swift}}
\newcommand{\N}{\mathcal{N}}

\begin{document}


\title{Gravitational waves or deconfined quarks: what causes the premature collapse of neutron stars born in short gamma-ray bursts?}

\author{Nikhil Sarin}
 \email{nikhil.sarin@monash.edu}
\author{Paul D. Lasky}%
\author{Gregory Ashton}
\affiliation{School of Physics and Astronomy, Monash University, Vic 3800, Australia}
\affiliation{OzGrav: The ARC Centre of Excellence for Gravitational Wave Discovery, Clayton VIC 3800, Australia}

\date{\today}

\begin{abstract}
We infer the collapse times of long-lived neutron stars into black holes using the X-ray afterglows of 18 short gamma-ray bursts. 
We then apply hierarchical inference to infer properties of the neutron star equation of state and dominant spin-down mechanism.
We measure the maximum non-rotating neutron star mass $\Mtov{} = 2.31 ^{+0.36}_{-0.21} M_{\odot}$ and constrain the fraction of remnants spinning down predominantly through gravitational-wave emission to $\eta = 0.69 ^{+0.21}_{-0.39}$ with $68 \%$ uncertainties.
In principle, this method can determine the difference between hadronic and quark equation of states.
In practice, however, the data is not yet informative with indications that these neutron stars do not have hadronic equation of states at the $1\sigma$ level.
These inferences all depend on the underlying progenitor mass distribution for short gamma-ray bursts produced by binary neutron star mergers.
The recently announced gravitational-wave detection of GW190425 suggests this underlying distribution is different from the locally-measured population of double neutron stars.
We show that $\Mtov{}$ and $\eta$ constraints depend on the fraction of binary mergers that form through a distribution consistent with the locally-measured population and a distribution that can explain GW190425. 
The more binaries that form from the latter distribution, the larger $\Mtov{}$ needs to be to satisfy the X-ray observations.
Our measurements above are marginalised over this unknown fraction. 
If instead, we assume GW190425 is not a binary neutron star merger, i.e the underlying mass distribution of double neutron stars is the same as observed locally, we measure $\Mtov{} = 2.26 ^{+0.31}_{-0.17} M_{\odot}$.
\end{abstract}

\pacs{Valid PACS appear here}
\maketitle

\section{\label{sec:intro}Introduction}
The historic first detection of gravitational waves from a binary neutron star inspiral GW170817, ushered in a new era of gravitational-wave and electromagnetic multi-messenger astronomy \citep{abbott17_gw170817_detection, abbott17_gw170817_multimessenger,GW170817_Integral,GW170817_FermiGRB_paper} and confirmed that binary neutron star mergers are the progenitors of some short gamma-ray bursts \citep[see e.g.,][]{abbott17_gw170817_multimessenger}.
Short gamma-ray bursts are typically accompanied by lower energy broadband emission, commonly attributed to the interaction of the jet with the surrounding interstellar medium \citep[e.g.,][]{meszaros93,piran98,sari99,granot99,meszaros01}. 
However, the X-ray afterglow of some short gamma-ray bursts often exhibits two features that cannot be adequately explained by such an interaction; a plateau and steep decay hundreds to thousands of seconds after the burst~\citep{zhang05,fan06,rowlinson10, rowlinson13, lu15}.
Although, there have been recent interpretations of sharp drops as a signature of a reverse shock \citep{beniamini17, lamb19}.
These features can be explained by a long-lived, rapidly-rotating, highly-magnetized neutron star~\cite[e.g.,][]{dai98,zhang01,lasky17,sarin19,xue19,xiao19,strang19}.
A steep decay is attributed to the collapse of such a neutron star into a black hole \citep{troja07, rowlinson10}. 
Such supramassive neutron stars collapse because they are born above the non-rotating neutron star mass limit---the Tolman-Oppenheimer-Volkoff mass $\Mtov{}$---but collapse when they lose the additional centrifugal support required to prevent black hole formation. This is different to hypermassive neutron stars which collapse on shorter timescales~\citep[e.g.,][]{lucca19}.
In contrast, the observations of a plateau with no sharp drop are best interpreted as the signature of a stable long-lived neutron star, possible if the neutron star is born with mass below $\Mtov{}$.

Several authors have attempted to indirectly infer the neutron star equation of state given the aforementioned plateau and steep decay features. 
This is done through inferring the ratio of short gamma-ray bursts that produce supramassive or infinitely stable neutron stars \citep[e.g.,][]{lu15}, or by measuring the collapse time which is a function of the equation of state, the dominant spin-down mechanism, and the progenitor mass distribution~\cite{lasky14,ravi14,li16_ang, gao16}.
The idea that the collapse time of these objects come from a distribution with the shorter than expected collapse time perhaps being an indication of gravitational-wave emission was suggested by \citet{fan13}.
\citet{ravi14} derived a theoretical collapse-time distribution assuming supramassive neutron stars spindown predominantly through magnetic-dipole radiation, finding the four reliable collapse-time measurements at that time to be smaller, and seemingly at odds with the theoretical distribution.
This discrepancy between the observed and theoretical distributions has been interpreted as evidence for two alternative hypotheses; the existence of deconfined quarks \citep{li16_ang, drago16,drago18} or initial rapid spin-down through gravitational waves \citep{fan13,gao16}.
The task of this paper is to determine which of these interpretations is correct.

We fit a collapsing neutron star model to the light-curves of all short gamma-ray burst X-ray afterglows observed by The Neil Gehrels \swift{} Telescope measuring the collapse time of $18$ neutron stars born in short gamma-ray bursts.
We perform Bayesian hierarchical inference to infer hyperparameters associated with the equation of state and dominant spin-down mechanism.
This involves first, measuring the collapse time from the X-ray afterglow then inferring the parameters associated with the distribution of collapse times.

We measure $\Mtov{} = 2.31 ^{+0.36}_{-0.21} M_{\odot}$; uncertainties here and throughout are $1\sigma$ unless otherwise stated. 
We constrain the fraction of neutron stars spinning down predominantly through gravitational-wave emission to $\eta = 0.69 ^{+0.21}_{-0.39}$, suggesting $\sim 70~\%$ of these neutron stars spin down predominantly~through gravitational waves. 
Although the gravitational waves emitted from these objects are likely not detectable individually, this constraint has important implications for the gravitational-wave stochastic background and the mechanisms which generate gravitational waves in these objects, such as the spin-flip or bar-mode instability.
We also measure equation-of-state specific parameters which indicates that the data is best explained by quark star equation of states at the $1\sigma$ level.

These results all depend on the underlying binary neutron star mass distribution, which has been typically assumed to be the same as the galactic double neutron star mass distribution observed locally with radio.
However, the gravitational-wave event GW190425 \citep{abbott19_190425_detection} is a massive binary neutron star merger with progenitor masses inconsistent with the local population.
With a total mass~$\sim 3.4 M_{\odot}$, GW190425 may have formed dynamically rather than through isolated binary evolution or perhaps through case-BB common envelope phase \citep{abbott19_190425_detection}.
Conservatively, this suggests the neutron star binaries that merge to produce short gamma-ray bursts are a mixture of the locally observed binary neutron star mass distribution and a mass distribution that can explain GW190425.

We perform our analysis with a modified mass distribution that allows for a bimodal distribution consistent with all neutron stars in our galaxy. 
We parameterize this distribution with an unknown mixing fraction dictating the probability of neutron stars coming from the two aforementioned formation channels.
Our results above are marginalised over this unknown mixing fraction.
If instead, we assume neutron star binaries that merge to produce short gamma-ray bursts are drawn equally from both distributions we measure $\Mtov{} = 2.30 ^{+0.38}_{-0.19} M_{\odot}$. 
If instead we assume a mixing fraction $\epsilon = 0$, i.e a distribution that can explain the progenitors of the locally observed binary neutron stars and GW170817 but one that cannot explain GW190425, then we measure $\Mtov{} = 2.26 ^{+0.31}_{-0.17} M_{\odot}$.

In this paper, we introduce our model for a collapsing magnetar and present the collapse-time probability distributions and lightcurves of $18$ short gamma-ray bursts in Sec.~\ref{sec:collapsetime}.
In Sec.~\ref{sec:methodology} we derive our Bayesian hierarchical model.
In Sec.~\ref{sec:Popinf} we show our results for the nuclear equation of state and spin-down mechanism and discuss the implications of our analysis. 
We discuss limitations and future extensions of our analysis and conclude in Sec.~\ref{sec:conclusion}.
\section{neutron star collapse times}\label{sec:collapsetime}
Rapidly rotating, millisecond magnetars were first introduced as an alternative central engine for gamma-ray bursts \citep{dai98, zhang01} and have been incredibly successful in interpreting the \swift{} X-ray afterglow observations of several short gamma-ray bursts \citep[e.g.,][]{fan06, rowlinson10, rowlinson13, lu15}.
The standard fireball-shock model governs the emission produced from the interaction of the jet with the surrounding interstellar medium. 
A model that has been modified in several ways to explain the plateau observations such as through the evolution of the microphysical parameters of the forward shock \citep{ioka06}, long-lived reverse shocks \citep{uhm07} and several other modifications \citep[e.g.,][]{toma06, oganesyan19}.
However, these modifications cannot adequately explain the steep decay feature which is naturally included in the magnetar model as the signature of a neutron star collapsing into a black hole \citep[e.g.,][]{rowlinson10}.

\citet{lasky14} derived a model for the collapse time assuming these newly-born neutron stars spin down only through vacuum dipole radiation, which has been used to model the collapse time of several candidate neutron stars born in short gamma-ray bursts \citep[e.g.,][]{lu16}. 
However, such modelling is fraught with difficulties with systematic uncertainties from k-corrections, restriction to modelling only for gamma-ray bursts with a measured redshift, and assumption of a vacuum dipole spin-down mechanism. 
The latter assumption is problematic as the braking index of two putative neutron stars born in GRB130603B and GRB140903A find only the former to be consistent with spindown through dipole radiation in vacuum.

The optimal approach is to directly measure the collapse time as the time of the sharp drop in the X-ray afterglow as done for GRB090515 \citep{rowlinson10} and then extended to a full catalogue of short gamma-ray bursts \citep{rowlinson13}. 
Here we do a similar analysis with the extended model from \citet{lasky17} that allows for spin-down through arbitrary braking indices as opposed to the model used by \citet{rowlinson13} which was restricted to spindown with a fixed braking index.
Our model for the luminosity evolution of a collapsing magnetar as derived in \citet{lasky17} is,
\begin{equation}\label{eqn:collapsing_mag}
L(t) = At^{\Gamma} + \mathcal{H}(t - \tcol)L_{0}\left(1 + \frac{t}{\tau}\right)^{\frac{1 + n}{1 - n}}.
\end{equation}
Here, $L$ is the luminosity, $t$ is the time since burst, $n$ is the braking index, $A$ and $\Gamma$ are the power-law amplitude and power-law exponent respectively, which together describe the emission from the tail of the prompt, $L_0$ is the initial luminosity at the onset of the plateau phase, $\tau$ is the spin-down timescale, and $\tcol{}$ is the collapse time. 
We note that since we fit to the flux data, the quantities here are in the detector frame and are later transformed into the source frame as we elaborate below.
The second term in Eq.~(\ref{eqn:collapsing_mag}) is the magnetar model from \citet{lasky17}, which models the luminosity evolution of a neutron star spinning down with an arbitrary braking index, with the step-function modification switching off this emission at a time $\tcol{}$.
We fit our model to all short gamma-ray bursts with X-ray afterglow data since the launch of \swift{} using the nested sampler \dynesty~\cite{dynesty} through the Bayesian inference library \bilby~\citep{bilby}. 
Our Priors on the various parameters are listed in Table.~\ref{table:priors}.
\begin{table}[t!]
\centering
\begin{tabular}{||c c||} 
 \hline
 Parameter & Prior \\ [0.5ex] 
 \hline\hline
 $A$ & $\log \textrm{Uniform}[10^{-20},10^{2}]$ \\ 
 $\Gamma$ & $\textrm{Uniform}[-4,-1]$ \\
 $L_{0}$ & $\log \textrm{Uniform}[10^{-20},10^{-9}]$ \\
 $\tau$ & $\log \textrm{Uniform}[10^{2},10^{7}]$ \\
 $\tcol{}$ & $\log \textrm{Uniform}[10^{1},10^{7}]$ \\
 $n$ & $\textrm{Uniform}[2,7]$ \\ [1ex] 
 \hline
\end{tabular}
\caption{Priors used to fit the collapsing magnetar model using Eq.~\ref{eqn:collapsing_mag}.}
\label{table:priors}
\end{table}

In contrast to \citet{rowlinson13} who assumed an average redshift for gamma-ray bursts without redshift information, we fit directly to the flux lightcurve.
Our inference allows us to measure the collapse time directly from the flux lightcurve which we then convert to the source frame by randomly drawing redshift samples from a probability distribution for $z$, $P(z)$.
For gamma-ray bursts with a known redshift, $P(z)$ is defined as a Gaussian around the known redshift with mean and standard deviation obtained from the \swift{} database, while for gamma-ray bursts where the redshift is not known, we define $P(z)$ as uniform in co-moving volume between a redshift $z = 0$ and $z = 2$. 
We obtained our flux data for all short gamma-ray bursts from the \swift{} database binned using the automatic binning strategies \citep{evans10_swiftdata}.

\begin{figure}[ht]
\centering
\includegraphics[width=0.5\textwidth]{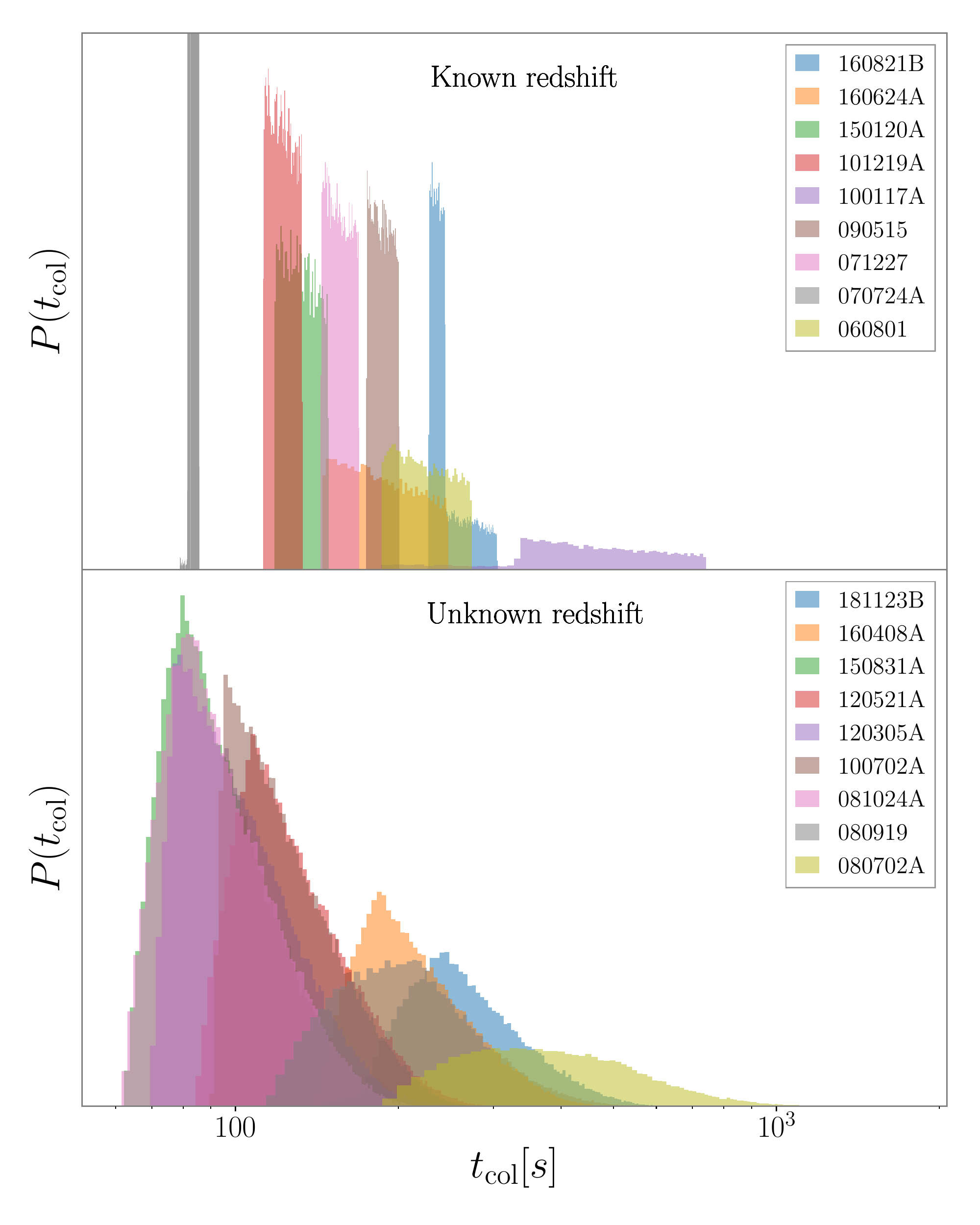}
\caption{One-dimensional posterior distributions for the collapse times of all short gamma-ray bursts that have observations supporting a collapsing neutron star model. The top panel shows posteriors for short gamma-ray bursts with known redshifts, while the bottom panel shows posteriors for gamma-ray bursts with unknown redshifts.}
  \label{fig.collapsetimes}
\end{figure}

We show our one-dimensional marginalized posterior for the source frame $\tcol{}$ in Fig.~\ref{fig.collapsetimes}, with the top panel showing collapse-time measurements for short gamma-ray bursts with known redshift measurements, while the bottom panel shows the collapse-time measurements for gamma-ray bursts without a measured redshift.
Our inference allows us to obtain posteriors for all six parameters for each gamma-ray burst.
An interesting feature of the posterior is the top-hat structure. 
This is a product of the uncertainty in measuring the collapse time as the time of the sharp drop in X-ray flux and limited to the resolution of the data, i.e., $\tcol{}$ could be anywhere between two data points where the sharp drop occurs.

\begin{figure*}
        \includegraphics[width=\textwidth,height=24cm]{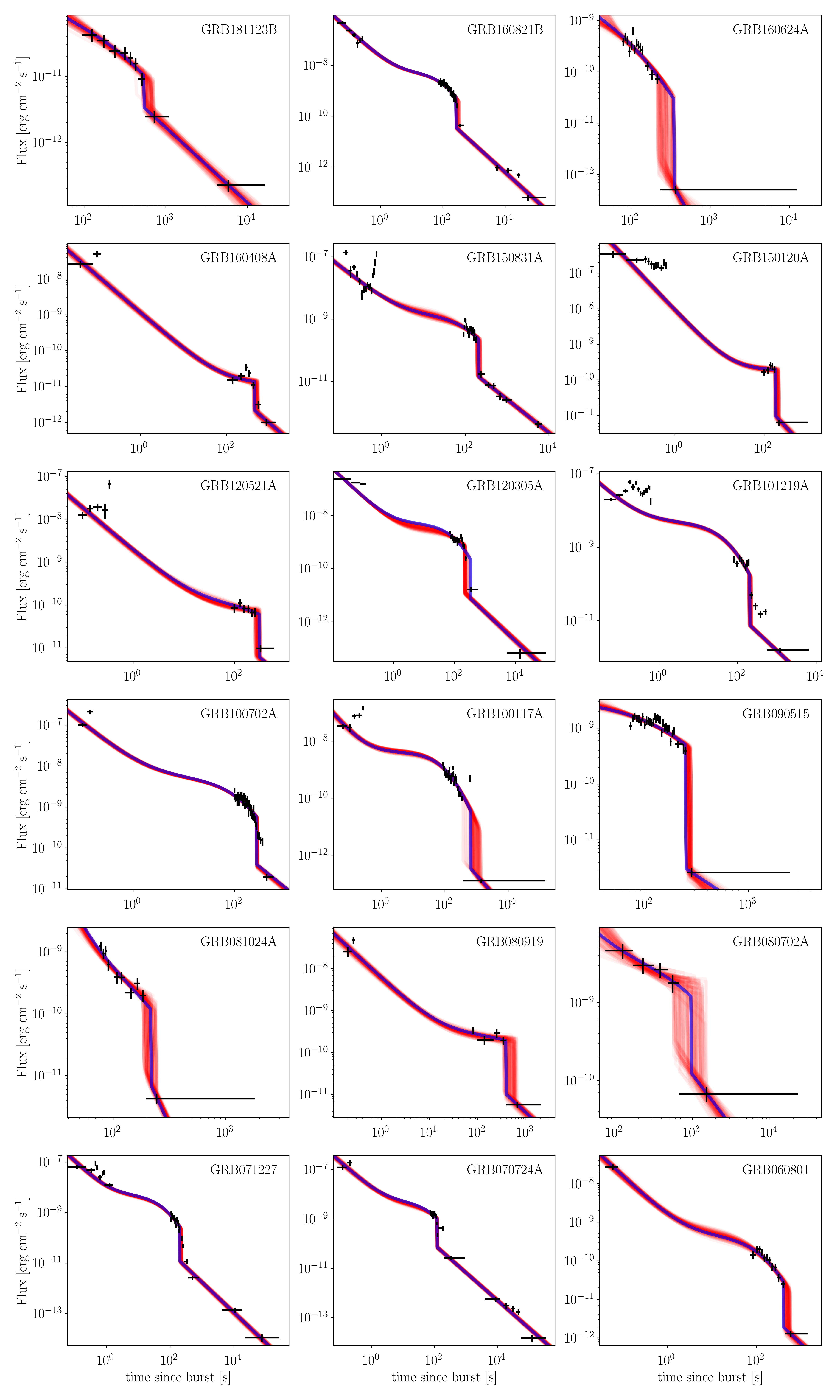}
  \caption{X-ray lightcurves for all gamma-ray bursts indicative of a collapsing neutron star. Black points indicate flux data from \textit{Swift} binned using the Swift automated binning strategy. The blue curve shows the maximum likelihood model for the collapsing magnetar model (Eq.~(\ref{eqn:collapsing_mag})). The dark red band is the superposition of $100$ predicted lightcurves randomly drawn from the posterior distribution.} 
\label{Fig. allcurves}  
\end{figure*}

Although a sharp drop in luminosity cannot be adequately explained within the fireball-shock model, we perform Bayesian model selection between our collapsing magnetar model and an agnostic fireball-shock model as described in \citep{sarin19} to ensure the data is best explained by a collapsing magnetar model. 
The Bayes factors comparing the fireball-shock and magnetar model for these $18$ gamma-ray bursts are shown in Table~\ref{table:bf_modelselection}. 
\begin{table}[ht!]
\centering
\begin{tabular}{||c c||} 
 \hline
 GRB & $\ln BF_{M/F}$ \\ [0.5ex] 
 \hline\hline
 GRB181123B & 12 \\ 
 GRB160821B & 1874 \\
 GRB160624A & 112 \\
 GRB160408A & 28 \\
 GRB150831A & 522 \\ 
 GRB150120A & 3 \\
 GRB120521A & 80 \\
 GRB120305A & 419 \\
 GRB101219A & 208 \\ 
 GRB100702A & 1752 \\
 GRB100117A & 756 \\
 GRB090515 & 732 \\
 GRB081024A & 37 \\ 
 GRB080919 & 53 \\
 GRB080702A & 6 \\
 GRB071227 & 430 \\
 GRB070724A & 362 \\ 
 GRB060801 & 162 \\ [1ex] 
 \hline
\end{tabular}
\caption{Bayes factor $\ln BF_{M/F}$ for the collapsing magnetar model introduced here (Eq.~\ref{eqn:collapsing_mag}) and fireball-shock model as introduced in \citep{sarin19}.}
\label{table:bf_modelselection}
\end{table}
As these Bayes factors indicate, assuming both models are equally likely\footnote{~In reality, both models are not equally likely as the fireball is always believed to be present.~Here, the correct metric to compare the two models is the Odds (see \citet{sarin19} for details), however model selection with the Odds requires knowing $\Mtov{}$ and the neutron star mass distribution.}, the collapsing magnetar model is significantly favoured over the fireball-shock model indicating that the X-ray afterglow observations here are best explained by the presence of a long-lived neutron star which collapses at some time.
Of the set of gamma-ray bursts considered, GRB150120A and GRB080702 have the lowest Bayes factors, albeit still positive indicating preference for the collapsing magnetar model.
To demonstrate our overall conclusions are not biased by these results, we repeat our hierarchical inference analysis without these two gamma-ray bursts and with leave-one-out cross validation: we find the same overall conclusions.
We show fits to all short gamma-ray burst X-ray afterglows that are best-fit by our model (Eq.~(\ref{eqn:collapsing_mag})) in Fig.~\ref{Fig. allcurves}.
\section{\label{sec:methodology}Methodology}
While individual collapse-time measurements are insightful, particularly if accompanied by the detection of gravitational waves from the binary neutron star inspiral \citep[e.g.,][]{lasky14}, significant constraints on the nuclear equation of state and spin-down mechanism can be placed by considering the population.
Hierarchical Bayesian inference is a formalism that can accurately measure population parameters. 
Here we write the formalism specifically for our problem; see \citet{mackay02} for a general discussion and derivation.

As discussed in Sec.~\ref{sec:intro}, there are two hypotheses in the literature to explain the inconsistency between the measured collapse times and the theoretical distribution \citep{ravi14}. 
However, as we noted in Sec.~\ref{sec:collapsetime}, the model for the collapse time used in literature is derived assuming the neutron star is spinning down solely through vacuum dipole radiation. 
We extend this model to include spindown via arbitrary braking indices through the general torque equation
\begin{equation}
\dot{\Omega}=k \Omega^{\nn{}}.
\end{equation}
Here, $\Omega$ is the star's angular frequency, $\dot{\Omega}$ is its time derivative, and $\nn{}$ is the \textit{averaged} braking index.
We emphasize that this \textit{averaged} braking index is different from the braking index measured through the fitting of Eq.~(\ref{eqn:collapsing_mag}) to the X-ray afterglow as the braking index there is measured at later times after the spin-down timescale $\tau$ as the braking index likely evolves as the dynamics of the newly-born neutron star change \citep[e.g.,][]{mus19}. 
One can see this more clearly by considering Fig.~\ref{Fig. allcurves}, given our model for the luminosity evolution (Eq.~\ref{eqn:collapsing_mag}), the measurement of $n$ comes after $t > \tau$, i.e., after the end of the plateau, as it dictates the shape of the power-law at the end of the plateau. 
The braking index is not measured earlier during the plateau, where it is quite likely different.

Using the general torque equation, one can derive a functional form of the evolution of the \textit{averaged} spin period as a function of time 
\begin{equation}\label{eqn:p(t)}
p(t)=p_{0}\left(1+\frac{t}{\tau}\right)^{\frac{\nn{} - 1}{\nn{}+1}}.
\end{equation}
Here, $p_0$ is the initial spin-period of the neutron star and $p(t)$ is the spin period as a function of time. 
The maximum gravitational mass, $M_{\max}$, of a spinning neutron star for a given equation of state can be written as \citep{shapiro83},
\begin{equation}\label{eqn:m_max}
M_{\max}=\Mtov{}\left(1+\alpha p^{\beta}\right)
\end{equation}
Here, $\alpha$ and $\beta$ are parameters fit to neutron star equilibrium sequences calculated for various values of the spin period, $p$. 
In Newtonian gravity, $\beta=-2$ and $\alpha$ is a function of the star's mass, radius and moment of inertia.
Together, $\alpha$ and $\beta$ describe an equation of state and have been calculated for several equations of state \citep[e.g.,][]{lasky14,li16_ang}.
To make our analysis cleaner, we nondimensionalize Eq.~(\ref{eqn:m_max}) by introducing a reference spin period, $p_{\rm{ref}}$
\begin{equation}\label{eqn:m_max_modified}
M_{\max}=\Mtov{}\left[1+\bar{\alpha} \left(\frac{p}{p_{\rm{ref}}}\right)^{\beta}\right],
\end{equation}
where $\bar{\alpha} = \alpha p_{\rm{ref}}^{\beta}$ is a dimensionless variable related to $\alpha$. 
Substituting Eq. (\ref{eqn:p(t)}) into Eq.~\ref{eqn:m_max_modified} and setting $M_{\max}$ to $M_p$ and $t$ to $\tcol$ gives 
\begin{equation}\label{eqn:generalcollapsetime}
\tcol{}_{,i} =\frac{\tau_{i}}{p_{0,i}^{\gamma_i}}\left[\left(\frac{M_{p,i}-\Mtov{}}{\alpha \Mtov{}}\right)^{\frac{\gamma_i}{\beta}}-p_{0,i}^{\gamma_i}\right].
\end{equation}
Here
\begin{equation}
    \gamma_i = \frac{\nn{}_{i}+1}{\nn{}_{i}-1},
\end{equation}
$M_p$ is the mass of the post-merger remnant, $p_{\rm{ref}}$ is a reference spin period which we set to $1$ ms without loss of generality.
Parameters denoted with $i$ are individual event parameters and those without are the population parameters we want to infer. 
Although $\Mtov{}$ can be calculated explicitly by determining $\alpha$ and $\beta$, the relationship is not unique and as such we have conservatively assumed that these parameters are uncorrelated.

Of the parameters denoted with $i$, we measure $\tau$ from the X-ray afterglow, albeit poorly if the neutron star collapses before $\tau$. 
Our initial parameter estimation on the X-ray afterglow also measures the braking index, $n$. 
However, as we emphasized above this $n$ is different to $\nn{}$. 
Instead, we model $\nn$ as either being indicative of predominant spin down through gravitational-wave emission or through an unknown braking index which we measure.
This implies that we model $\nn{}$ to be randomly drawn from the distribution described by 
\begin{equation}\label{eqn:npdf}
\nn = (1 - \eta)\mathcal{N}(\mu_{\nn{}},\sigma_{1}) + \eta\mathcal{N}(5, \sigma_{2}),
\end{equation}
where $\mathcal{N}\left(\mu,\,\sigma\right)$ is a Gaussian distribution of mean $\mu$ and standard deviation $\sigma$, $\eta$ is a mixing fraction between the two Gaussian distributions, $\mu_{\nn}$ is the mean of the first Gaussian distribution and $\sigma_{1}$ and $\sigma_{2}$ are the standard deviations of the first and second Gaussian distributions.
This implies that the population of \textit{average} braking index is a mixture model of two Gaussian distributions, one centred on $\nn = 5$ implying an \textit{average} braking index where the spin down of the neutron star is dominated by gravitational-wave emission and another Gaussian distribution centred on $\mu_{\nn}$ which we infer.
We emphasize that this model is a choice and we believe it captures the necessary physics. 

Equations~(\ref{eqn:generalcollapsetime}) and~(\ref{eqn:npdf}) together describe our population model, parameterized by hyperparameters, $\Lambda = \{\bar{\alpha}, \beta, \Mtov, \mu_{\nn}, \sigma_{1}, \sigma_{2}, \eta\}$.
By Bayes' theorem the posterior distribution on these hyperparameters is 
\begin{equation}\label{eqn:bayes}
p_{\mathrm{tot}}(\Lambda | \vec{d})=\frac{\mathcal{L}_{\mathrm{tot}}(\vec{d} | \Lambda) \pi(\Lambda)}{\int d \Lambda \mathcal{L}_{\mathrm{tot}}(\vec{d} | \Lambda) \pi(\Lambda)}.
\end{equation}
Here, $\vec{d}$ is the set of measurements of $N$ events, $\pi(\Lambda)$ is our prior on the hyperparameters, and $\mathcal{L_{\mathrm{tot}}}(\vec{d}|\Lambda)$ is the likelihood of the population data given our hyperparameters. 
The denominator is the hyper-evidence, which can be used for comparing two population models. 
Naively, looking at Eq.~(\ref{eqn:bayes}) we might not see any dependence of our posterior on the event parameters. 
This relationship can be made explicit by rewriting the likelihood as
\begin{equation}\label{eqn:fulllikelihood}
\mathcal{L}_{\mathrm{tot}}(\vec{d} | \Lambda)=\prod_{i}^{N} \int d \theta_{i} \mathcal{L}\left(d_{i} | \theta_{i}\right) \pi\left(\theta_{i} | \Lambda\right).
\end{equation}
Here, $\theta_{i}$ is a vector of the $i^{\rm{th}}$ event parameters ($\theta_{i} = \{A, \Gamma, L_0, \tau, \tcol{}, n, M_{p}, p_{0}\}$), $d_{i}$ is the data for the $i^{\rm{th}}$ event, $\mathcal{L}\left(d_{i} | \theta_{i}\right)$ is the likelihood of the data $d_{i}$ given event parameters $\theta_{i}$ and $\pi\left(\theta_{i} | \Lambda\right)$ is the prior on $\theta_{i}$ given our hyperparameters.
These large sets of integrals in evaluating the hyper-likelihood make hierarchical inference prohibitively expensive, fortunately, a computational trick, referred to as ``recycling'' \citep[e.g.,][]{thrane19} replaces these integrals with sums over posterior samples from the initial step of parameter estimation on an individual event, in our case, the fitting of Eq. (\ref{eqn:collapsing_mag}) to the X-ray afterglow.

Our formulation is still not complete as there are two event-specific parameters we do not measure when fitting Eq.~(\ref{eqn:collapsing_mag}) to the X-ray afterglow, the mass of the post-merger remnant $M_p$, and the initial spin-period, $p_{0}$. 
We therefore marginalize over these two parameters, which can be written explicitly as
\begin{multline}\label{eqn:marginalisedlikelihood}
\mathcal{L}_{\mathrm{tot}}(\vec{d} | \Lambda)=\prod_{i}^{N} \iiint d\theta_{i}dM_{p,i}dp_{0,i} \mathcal{L}\left(d_{i} | \theta_{i}\right) \times \\
\pi\left(\theta_{i} | \Lambda\right) \pi\left(M_{p}|\Lambda\right)\pi\left(p_0|\Lambda\right),
\end{multline}
where $\pi\left(M_{p}|\Lambda\right)$ and $\pi\left(p_0|\Lambda\right)$ are the prior distributions on $M_{p}$ and $p_0$ given our hyperparameters.
We assume a uniform prior on $p_0$ from $0.5-1.0$ ms, although we note that in reality the spin-period prior should be a function of the hyperparameters, in particular, $\alpha$ and $\beta$.
However, given we are marginalising over this parameter, we have conservatively accounted for this covariance by propagating all of the uncertainty through to our measured parameters.

The prior on the post-merger remnant mass distribution, $\pi\left(M_{p}|\Lambda\right)$ is much more complicated.
Previously, several authors have calculated the distribution of $M_{p}$ using the observed binary neutron star population in our galaxy \citep{lasky14,lu15,sarin19}. 
These galactic double neutron star systems measured with radio observations are empirically known to have a tight mass distribution described by a Gaussian of mean $\mu=1.32M_\odot$ and width $\sigma=0.11M_\odot$~\cite{kiziltan13,alsing18}.
While the progenitors of GW170817 are consistent with the galactic double neutron star mass distribution~\cite{farrow19}, the progenitors of GW190425 are not at a highly-significant level~\cite{abbott19_190425_detection}.  
This suggests GW190425 came from a different population, perhaps as a result of dynamical formation or unstable case-BB common-envelope evolution \citep{abbott19_190425_detection}.
In this case, one would expect the masses of the progenitors of GW190425 to be drawn from the population of neutron stars \textit{not} in double neutron star systems and instead from a mass distribution consisting of all neutron stars.

Following \citet{alsing18}, we use the galactic neutron star mass distribution, consisting both populations of double neutron stars and neutron stars in other systems, to be representative of the underlying mass distribution of progenitors for gravitational-wave mergers and short gamma-ray bursts.  
If GW190425 originated through a different evolutionary pathway than observed galactic double neutron star systems, then the relative fraction between the two populations is almost entirely unknown. 
Indeed, while it was originally thought the progenitor of GW170817 came from the same population as galactic double neutron stars~\cite[e.g.,][]{farrow19}, even this should now be called into question.
As a consequence, we leave the mixing fraction between the two populations as a free parameter.
The full population of galactic neutron stars can be fit with a double-peaked Gaussian probability distribution~\cite{alsing18}
\begin{align}
    p(M)=\left(1-\epsilon\right)\N\left(\mu_1,\,\sigma_1\right)+\epsilon\N\left(\mu_2,\,\sigma_2\right),\label{eq:pm}
\end{align}
The known galactic systems have $\mu_1=1.32M_\odot$ and $\sigma_1 = 0.11$, $\mu_2=1.80M_\odot$, $\sigma_2=0.21M_\odot$, and mixing fraction $\epsilon=0.35$.

\begin{figure*}
    \centering
    \includegraphics[width=0.95\textwidth]{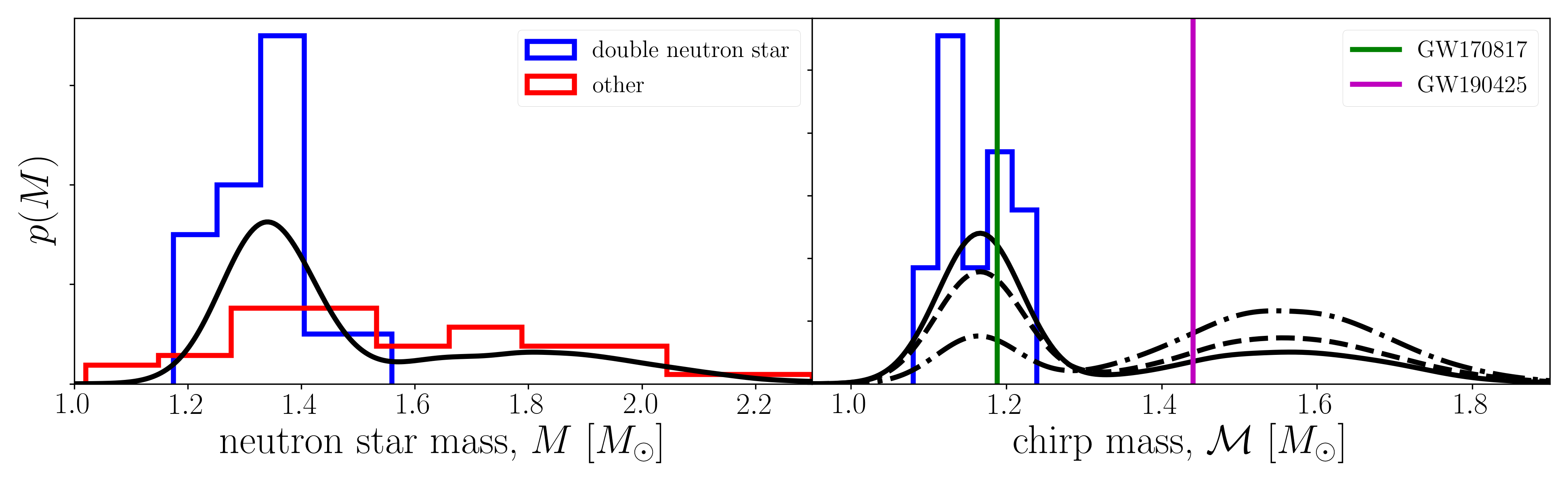}
    \caption{Neutron star mass distributions.  Left panel: In blue are the measured neutron star masses for those in double neutron star systems, and in red are the masses of neutron stars in binaries with white dwarfs, main sequence stars, etc~\cite{alsing18}.  The black curve is the best-fit mass distribution to these from~\citet{alsing18}. Right panel: in blue are the same double neutron star systems, this time converted to chirp mass.  In green and magenta are the chirp masses of the two gravitational-wave events GW170817 and GW190425, respectively.  The solid black curve is the chirp-mass distribution associated with the solid black curve in the left panel.  The dashed and dot-dashed black curves assume similar distributions to the solid-black curve, except the mixing fraction between the two binary populations is $\epsilon=0.5$ and $0.8$, respectively (cf. $\epsilon=0.35$ for the solid black curve).}
    \label{fig:BNSmasses}
\end{figure*}

In the left panel of Fig.~\ref{fig:BNSmasses} we show these mass distributions. 
The blue histogram shows the masses of neutron stars in galactic double neutron star systems, while the red histogram shows the masses of all other neutron stars.
In black is the probability distribution given by Eq.~(\ref{eq:pm}) with values given above. 
In the right-hand panel of Fig.~\ref{fig:BNSmasses}, we show in blue the corresponding histogram for the chirp masses of galactic double neutron stars; i.e., equivalent to the systems shown in blue in the left panel.  
The two vertical lines are the measured chirp masses for GW170817 in green and GW190425 in magenta; the uncertainties on these measurements are too small to be seen on this scale.
The solid black curve shows the chirp-mass probability distribution corresponding to converting the probability distribution of Eq.~(\ref{eq:pm}) into chirp mass.  The black dashed and dot-dashed curves show the same probability distribution, albeit with $\epsilon=0.5$ and $\epsilon=0.8$, respectively.

Inspecting Fig.~\ref{fig:BNSmasses} one can see that although the progenitors of GW190425 are inconsistent with the mass distribution inferred from galactic double neutron star systems, they are consistent with the mass distribution for all galactic neutron stars.
Assuming our galaxy is typical, one, therefore, expects the progenitor mass distribution for all binary neutron star mergers to be similar to the distribution given by Eq.~(\ref{eq:pm}), albeit with an unknown mixing fraction $\epsilon$.

For the remainder of the paper, we assume the progenitor mass distribution is given by Eq.~(\ref{eq:pm}) with $(\mu_1,\,\sigma_1)=(1.32,\,0.11)$,  $(\mu_2,\,\sigma_2)=(1.80,\,0.21)$, and let $\epsilon$ be a free parameter which we infer through our hierarchical model.
Following \citet{sarin19}, one can derive the post-merger remnant mass distribution having the same functional form as
Eq.~(\ref{eq:pm}) with $(\mu_1,\,\sigma_1)=(2.42,\,0.09)$,  $(\mu_2,\,\sigma_2)=(3.21,\,0.25)$ assuming $\approx 0.07 M_{\odot}$ of dynamical ejecta is produced in the merger, consistent with observations of GW170817 \citep[e.g.,][]{evans17}.

For the timescales we are interested in, neutron stars can only collapse if they are born with mass between $\Mtov{}$ and approximately $1.2 \times \Mtov{}$ implying
\begin{align}\label{eqn:priormp}
\pi\left(M_{p}|\Lambda\right) &= 
\begin{cases}
f\left(\epsilon\right)&\Mtov{}\leq M_{p} \leq 1.2\Mtov{} \\ 
0 & \text{otherwise}
\end{cases},
\end{align}
where $f\left(\epsilon\right) = \left(1-\epsilon\right)\N\left(2.42,0.09\right) + +\epsilon\N\left(3.21,0.25\right)$.
Our hierarchical likelihood is completely defined by Eqs.~(\ref{eqn:marginalisedlikelihood}-\ref{eqn:priormp}), ready to be combined with suitable priors on our hierarchical model (Eq.~\ref{eqn:generalcollapsetime}).
We perform hierarchical inference on our population of events using the nested sampler \dynesty~\cite{dynesty} through the Bayesian inference library \bilby~\citep{bilby}.
To make the analysis computationally feasible, we use an adaptation of the GPU-accelerated population inference code \gwpop~\citep{talbot19} and {\sc cupy}~\citep{cupy}. Our priors for the rest of the hyperparameters are shown in Table.~\ref{table:hyperpriors}.
\begin{table}[t!]
\centering
\begin{tabular}{||c c||} 
 \hline
 Parameter & Prior \\ [0.5ex] 
 \hline\hline
 $\Mtov{}$ & $\textrm{Uniform}[2.01,2.9]$ \\ 
 $\log_{10}\bar{\alpha}$ & $\textrm{Uniform}[-3,1]$ \\
 $\beta$ & $\textrm{Uniform}[-6,-2]$ \\
 $\sigma$ & $\textrm{Uniform}[1,500]$ \\
 $\mu_{\nn}$ & $\textrm{Uniform}[1,4]$ \\
 $\nn_{\sigma,1}$ & $\textrm{Uniform}[0.1,1.5]$ \\
 $\nn_{\sigma,2}$ & $\textrm{Uniform}[0.1,1.5]$ \\
 $\eta$ & $\textrm{Uniform}[0,1]$ \\ 
 $\epsilon$ & $\textrm{Uniform}[0,1]$ \\  [1ex] 
 \hline
\end{tabular}
\caption{Priors for our hierarchical model described by Eqs.~(\ref{eqn:marginalisedlikelihood}-\ref{eqn:priormp}).}
\label{table:hyperpriors}
\end{table}
\section{\label{sec:Popinf}Equation of state and gravitational-wave constraints}
\begin{figure}[ht]
\centering
\includegraphics[width=0.5\textwidth]{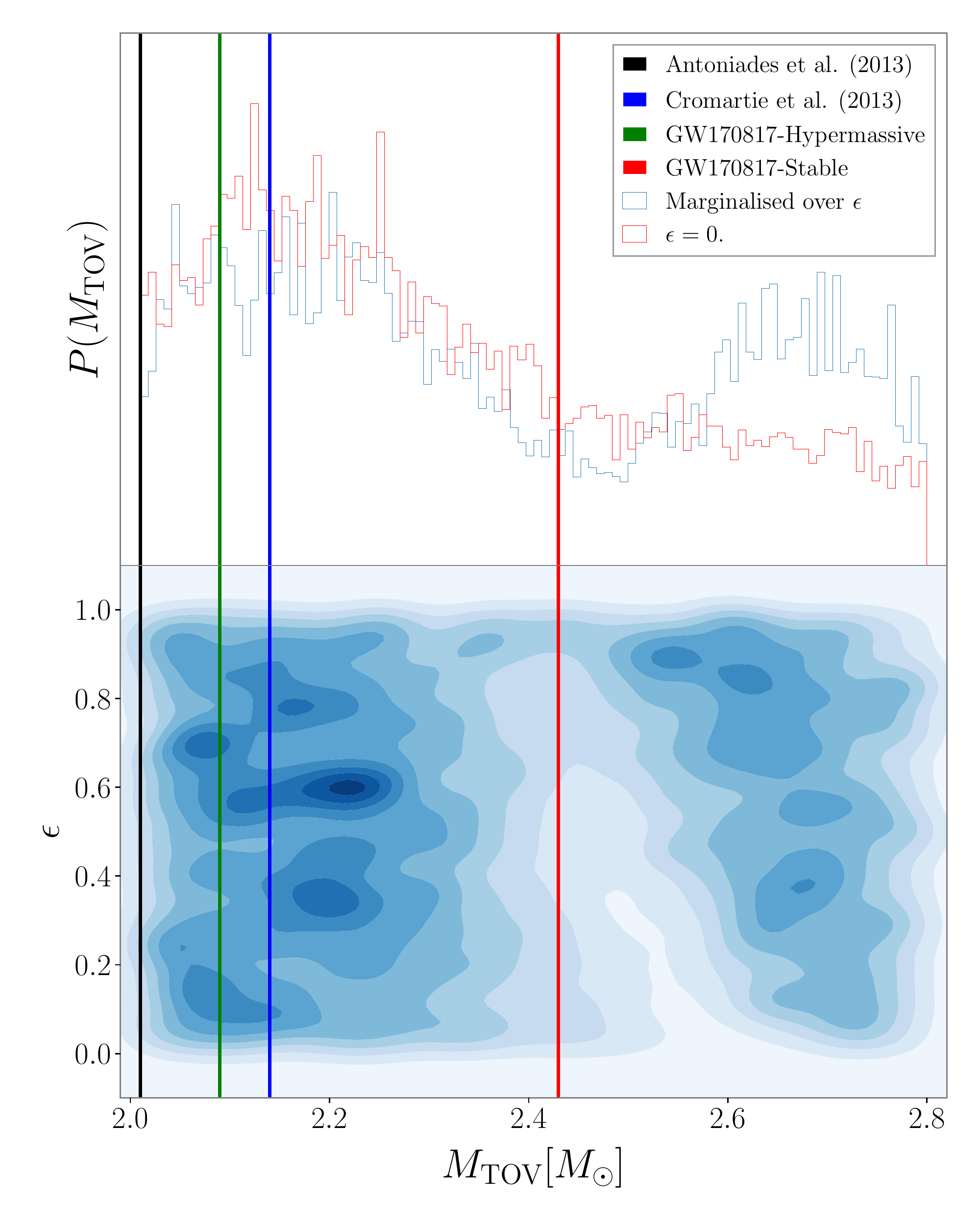}
  \caption{One (top panel) and two-dimensional (bottom panel) posterior distributions on $\Mtov{}$ and $\Mtov{}-\epsilon$. We also show a slice through the two-dimensional posterior for $\epsilon = 0.$ i.e a mass distribution similar to the galactic double neutron star systems but inconsistent with the progenitors of GW190425. We measure $\Mtov{} = 2.26 ^{+0.31}_{-0.17} M_{\odot}$ assuming a mixing fraction $\epsilon = 0.$ which implies a mass distribution inconsistent with the progenitors of GW190425.
  We plot few other constraints for $\Mtov{}$ based on pulsar observations \citep{antoniadis13, cromartie19} and inferred fate of GW170817 \citep[e.g.,][]{ai19}. For clarity, we only plot the median of these measurements but we stress that several of these measurements have large uncertainties and the later constraint, based on the inferred fate of the post-merger remnant of GW170817 could realistically be anywhere between the two hypermassive (green) or stable (red) scenarios.}
  \label{fig:mtov}
\end{figure}

We first show our measurement on the maximum allowed non-rotating mass $\Mtov{}$, as alluded to previously, this is a function of the unknown mixing fraction $\epsilon$, between double neutron stars observed in our galaxy and the population that explains the progenitors of GW190425. 
Our measurement for $\Mtov{}$ for mixing fraction, $\epsilon=0$, and marginalised over all possible values of this mixing fraction are shown in Fig.~\ref{fig:mtov} in the top panel. 
The bottom panel shows the two-dimensional posterior on $\Mtov{}$ and $\epsilon$.
On the same plot, we plot vertical lines for different constraints on $\Mtov{}$.
The black and blue lines correspond to the mass measurements of two pulsars, PSRJ0348+0432 and PSRJ0740+6620 as $2.01\pm 0.04M_{\odot}$ \citep{antoniadis13} and $2.14\pm 0.1M_{\odot}$ \citep{cromartie19} respectively, the existence of such massive neutron stars puts a lower limit on $\Mtov{}$.
The other two vertical lines come from the observation of GW170817, in particular by combining the mass measurement from the gravitational-wave inspiral and by inferring fate of the post-merger remnant.
However, there is still disagreement on the ultimate fate of the post-merger remnant of GW170817 with the interpretations of the electromagnetic observations ranging from a short-lived neutron star through to an infinitely stable neutron star. 
Such uncertainty on the fate of the post-merger remnant results in the constraint on $\Mtov{}$ ranging from $2.09-2.43 M_{\odot}$ \citep{ai19}.
The green and red vertical lines correspond to the limits of this constraint.
We note that the most widely accepted interpretation of the fate of the post-merger remnant of GW170817, a hypermassive neutron star that collapsed within $1.7$ seconds into a black hole constrains $\Mtov{} \lesssim 2.3M_{\odot}$ \citep[e.g.,][]{margalit17,shibata19}.

Assuming a mixing fraction $\epsilon = 0$, i.e a population consistent with local double neutron star systems but inconsistent with GW190425, we measure $\Mtov{} = 2.26 ^{+0.31}_{-0.17} M_{\odot}$. 
As Fig.~\ref{fig:mtov} shows, this is the most conservative measurement and comparable to other analyses measuring $\Mtov{}$ \citep[e.g.,][]{lu15, gao16} which assume the local binary neutron star population is a good representation of the binary neutron stars that merge. 
However, this mass distribution is inconsistent with GW190425. 
If instead we assume~$\epsilon=0.5$ we measure $\Mtov{} = 2.30 ^{+0.38}_{-0.19} M_{\odot}$.
We stress that with only two gravitational-wave observations of binary neutron star inspirals, it is impossible to constrain this mixing fraction. 
Instead, marginalizing over this unknown mixing fraction leads to $\Mtov{} = 2.31 ^{+0.36}_{-0.21} M_{\odot}$. 
We can revisit this measurement after future gravitational-wave measurements constrain $\epsilon$, allowing us to take a slice through our two-dimensional posterior for a fixed $\epsilon$.

We also measure the braking index mixing fraction $\eta = 0.69 ^{+0.21}_{-0.39}$ which suggests that $\sim 70\%$ of neutron star post-merger remnants that collapse spin down predominantly through gravitational-wave emission.
This has several consequences. 
Firstly, it is good for the prospect of detecting gravitational waves from these objects.
Although not individually resolvable with current detectors and even future detectors unless sufficiently close \citep[e.g.,][]{sarin18}, they will contribute to the stochastic background \citep[e.g.,][]{regimbau06,cheng17}, which may become detectable with third-generation gravitational-wave detectors \citep{cheng17}.
We leave a calculation of the stochastic background for future work.

The fraction of remnants that spin down through gravitational-wave emission is also interesting for understanding the emission mechanism itself.
It is intriguing to understand the physical difference between those remnants that do and do not spin down predominantly through gravitational-wave emission.
For example, there are a number of physical mechanisms that cause large-amplitude gravitational waves such as the spin-flip instability~\citep{cutler02}, inertial $r$ modes~\citep[e.g.,][]{andersson01}, or the secular bar-mode instability~\citep[see e.g.,][]{andersson03}.
Whether each of these mechanisms operate in certain remnants but not others could be a result of different initial conditions such as the progenitor masses.

The spin-flip instability in newly born neutron stars may operate when the internal toroidal magnetic field winds up, causing the star to become a prolate spheroid. 
Internal dissipation then causes the star to become an orthogonal rotator in which the dominant moment of inertia axis is misaligned with the star's rotation axis. 
In this configuration, the star is a maximal emitter of gravitational waves.
However, the birth magnetic field, temperature distribution, initial spin period can all play a large role in whether the spin-flip instability occurs or not~\cite[e.g., see][]{lander18}. 
Moreover, the spin-flip instability can cause the star to initially become an orthogonal rotator, before re-aligning and becoming an aligned rotator. 
In such a situation, one would expect significant gravitational-wave emission early in the star's life which then gets suppressed significantly as the star again becomes aligned~\citep[][]{dallosso18,lander19}.

Both the secular bar mode and inertial $r$ mode saturation amplitudes are highly uncertain, and likely depend on the star's temperature through bulk viscosity.
For example, if the star does not cool sufficiently ($10^{10}$~K), the bar-mode instability may be suppressed~\citep[e.g.,][]{doneva15} leading to a dearth of gravitational-wave emission.
The secular bar-mode instability might also fail if the ratio of $T/W$, where $T$ is the rotational kinetic energy and $W$ is the gravitational potential energy, simply does not exceed the critical point for the instability due to, for example, the mass ratio of the merging neutron stars.
While it is not clear what the active or dominant gravitational-wave emission mechanisms are in these nascent stars, it is clear that understanding the fraction that spin down through gravitational waves versus electromagnetic radiation could provide valuable insight into this interesting question.

\begin{figure}[ht]
\centering
\includegraphics[width=0.5\textwidth]{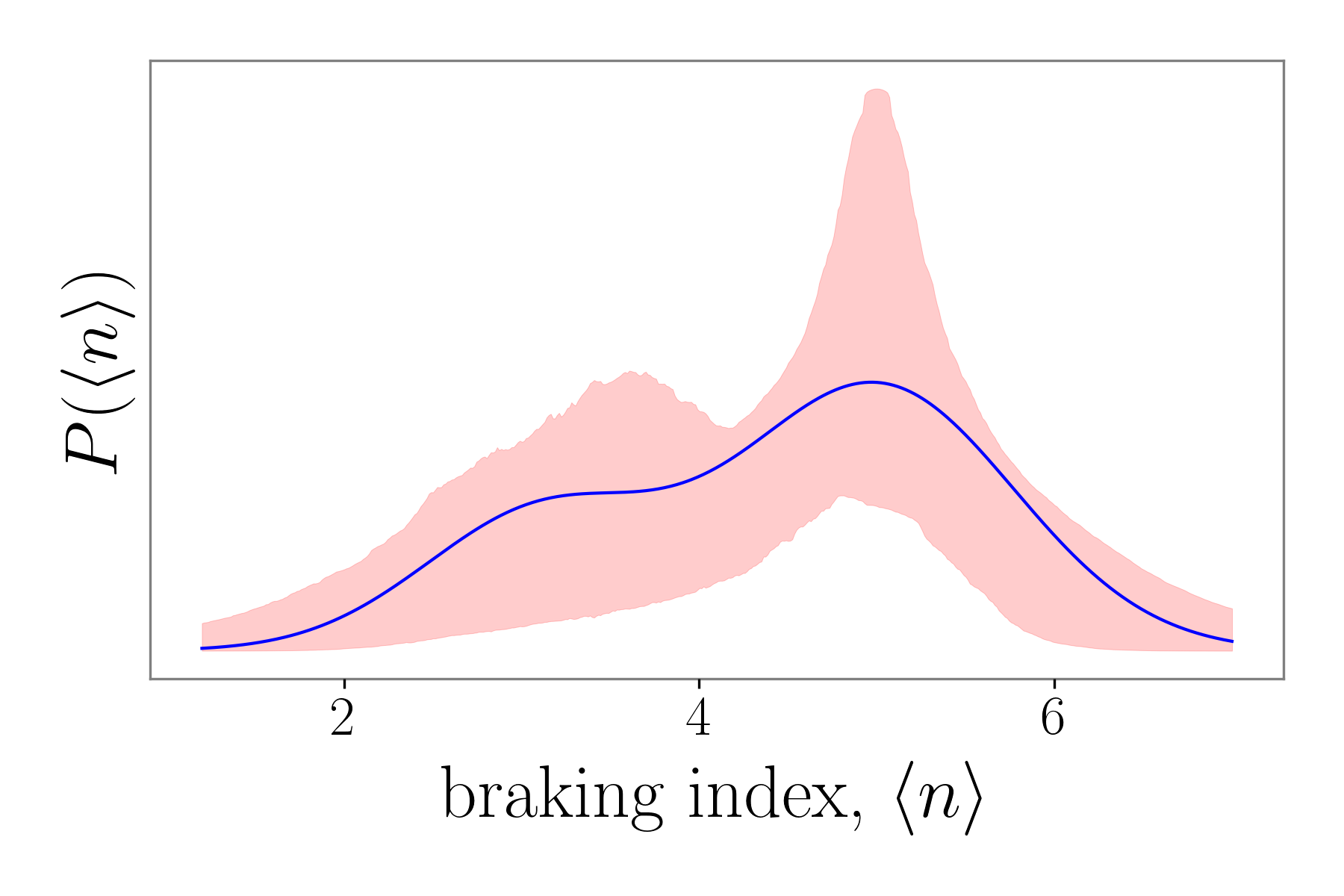}
  \caption{\textit{Average} braking index distribution. The blue curve indicates the median value of the posterior while the red curves are two-sigma confidence intervals.}
  \label{fig:braking_index}
\end{figure}

We measure $\mu_{\nn{}} = 3.12 ^{+0.69}_{-0.87}$ suggesting that the rest of the post-merger remnants that collapse spin-down through \textit{on average} close to vacuum dipole radiation.
Our measurement $\mu_{\nn} \gtrsim 3$ could imply we are seeing a mixture of gravitational-wave and electromagnetic emission, i.e., while $\sim 70\%$ are consistent with $\nn = 5$, the rest initially spin down through gravitational waves and later spin down through electromagnetic radiation.
We use our posteriors to construct the probability density function for the \textit{averaged} braking index which is shown in Fig. \ref{fig:braking_index}. 
This suggests that a large fraction of post-merger remnants that collapse spin-down predominantly through gravitational waves while the rest spin-down with an \textit{average} braking index close to $\nn{} = 3$, consistent with vacuum dipole radiation. 
We show the full two-dimensional posterior distribution on all these parameters in the Appendix~\ref{appendix}.

\begin{figure}[ht]
\centering
\includegraphics[width=0.5\textwidth]{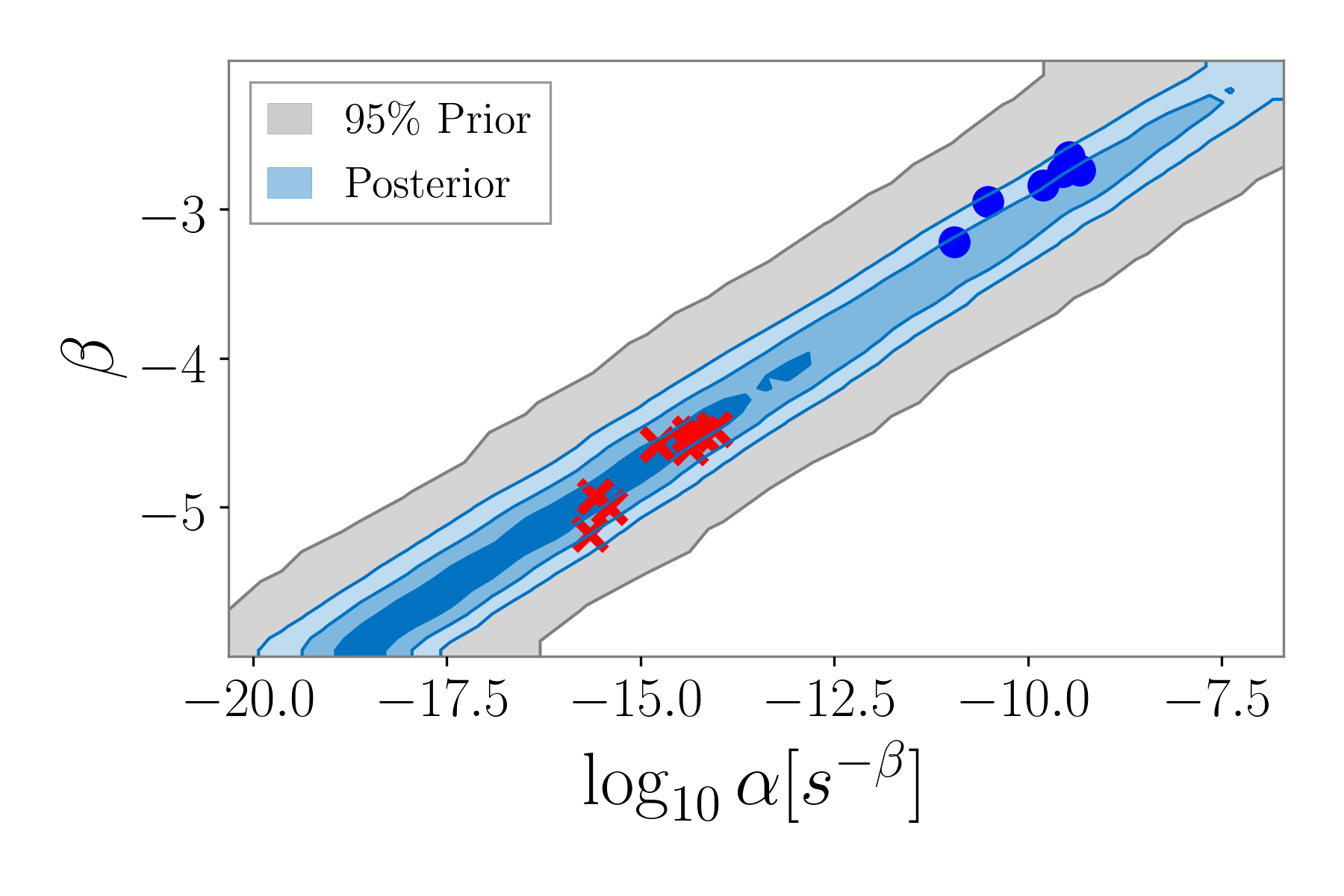}
  \caption{Two-dimensional posterior distribution of $\alpha$ and $\beta$ with hadronic equation of states marked by circles and quark star equation of states marked by crosses. The shades of blue correspond to one-two-three sigma confidence intervals.}
  \label{fig:alpha_beta}
\end{figure}

In Fig.~\ref{fig:alpha_beta} we show the two-dimensional posterior distribution of $\alpha$ and $\beta$, see Eq.~(\ref{eqn:m_max}). 
Here, hadronic equation of states are marked with blue dots while quark star equation of states are marked with red crosses.
The shade of blue in the posterior indicates the confidence level of our posterior and grey is the 95\% prior.
Our posterior is consistent with both quark and hadronic equations of state at the two-sigma level, with current constraints slightly favouring quark-like equations of states over purely hadronic.
The specific equation of states as well as their corresponding $\alpha$, $\beta$ parameters are listed in Table~\ref{table:eosalphabeta} but we emphasise that given the current size of the population we are not interested in individual equation of states, but rather the large difference in $\alpha-\beta$ parameter space between quark and hadronic equation of states.
The relationship between $\alpha$ and $\beta$ has been explored in the past with~\citet{ai19} exploring the constraints on these parameters for different equation of states with observations of GW170817 and \citet{gao20} deriving a general relationship for $\alpha$ and $\beta$ by parameterising in terms of how much more mass can be added for a given spin period.
\begin{table}[t!]
\centering
\begin{tabular}{||c c c||} 
 \hline
 Equation of state & $\alpha$ & $\beta$ \\ [0.5ex] 
 \hline\hline
 GM1 & $1.58\times10^{-10}p^{-\beta}$ & $-2.84$ \\
 APR & $0.303\times10^{-10}p^{-\beta}$ & $-2.95$ \\
 BSk20 & $3.39\times10^{-10}p^{-\beta}$ & $-2.68$\\
 BSk21 & $2.81\times10^{-10}p^{-\beta}$ & $-2.75$\\
 CIDDM & $2.58\times10^{-16}p^{-\beta}$ & $-4.93$\\
 CDDM1 & $3.938\times10^{-16}p^{-\beta}$ & $-5.0$\\
 CDDM2 & $2.22\times10^{-16}p^{-\beta}$ & $-5.18$\\
 MIT2 & $1.67\times10^{-15}p^{-\beta}$ & $-4.58$\\
 MIT3 & $3.35\times10^{-15}p^{-\beta}$ & $-4.60$ \\
 PMQS1 & $4.39\times10^{-15}p^{-\beta}$ & $-4.51$ \\
 PMQS2 & $5.90\times10^{-15}p^{-\beta}$ & $-4.51$ \\
 PMQS3 & $9.00\times10^{-15}p^{-\beta}$ & $-4.48$ \\ [1ex] 
 \hline
\end{tabular}
\caption{Equation of states and their corresponding $\alpha$ and $\beta$ parameters, all equation of states parameters are from \citet{li16_ang}.}
\label{table:eosalphabeta}
\end{table}
If these supramassive neutron stars are quark stars, this might suggest that either these newly-born neutron stars are born via the merger of two quark stars, or that the merger of two hadronic neutron stars results in a phase transition from a hadronic to quark equation of state. 
Both of these options have implications for nuclear theory, with the latter phase transition being perhaps detectable in the near-future with aLIGO \citep[e.g.,][]{chatziioannou19}.

In the future, with more events and a better informed binary neutron star mass distribution we will revisit these measurements.
\section{Conclusion} \label{sec:conclusion}
We have shown how observations of the X-ray afterglows of short gamma-ray bursts can be used to infer the presence of long-lived binary neutron star post-merger remnants that later collapse to form black holes.
This implicitly requires that long-lived neutron star post-merger remnants can produce a short gamma-ray burst, which is still heavily debated see \citep[e.g.,][]{giacomazzo13,Beniamini20,ciolfi20}
We have also shown that, given a population of these putative collapsing neutron stars, hierarchical Bayesian inference provides a framework for measuring the population properties. 
In particular, we use the observations of $18$ short gamma-ray bursts to measure the maximum allowed non-rotating mass $\Mtov{} = 2.31 ^{+0.36}_{-0.21} M_{\odot}$ marginalised over the unknown mixing fraction
between the mass distribution describing both single and double neutron stars observed in our galaxy, the former being consistent with the progenitors of GW190425. 
If instead, we assume $\epsilon = 0$ (i.e., a mass distribution that is inconsistent with the progenitors of GW190425 but a good representation of locally observed double neutron star systems), we measure $\Mtov{} = 2.26 ^{+0.31}_{-0.17} M_{\odot}$. 
Future measurements of gravitational waves from binary neutron stars will allow an independent measurement of $\epsilon$ allowing us to revisit our measurement and therefore provide a tighter constraint on $\Mtov{}$.

Although broad, our measurement for $\Mtov{}$ marginalised over the unknown mixing fraction is comparable to inferences of $\Mtov{}$ made with short gamma-ray bursts \citep[e.g.,][]{lu15}.
However, such measurements will need to be revisited as they assume the galactic double neutron star distribution is a good representation of binary neutron star merger progenitors. 
The observation of GW190425 suggests this is not the case.
Our measurement is also comparable to inferences of a sharp cut-off in the galactic neutron star mass distribution \citep{alsing18} and inference based on the uncertain nature of the post-merger remnant of GW170817 \citep{ai19}.

We measure equation-of-state specific parameters, $\log_{10}\alpha = -14.89^{+3.94}_{-2.72} \text{s}^{-\beta}$ and $\beta = -4.67^{+1.32}_{-0.92}$.
Together these measurements suggest deconfined quark equation of states are slightly favoured over hadronic, however, the data is not conclusive with both sets of equations of states being consistent with the population at the two-sigma level. 

We also measure the fraction of post-merger remnants that spin-down through gravitational waves implying a braking index, $n = 5$ as $\eta = 0.69 ^{+0.21}_{-0.39}$, suggesting that $\sim 70\%$ of neutron star post-merger remnants born in short gamma-ray bursts which collapse do so due to spin down predominantly through the emission of gravitational waves.

There are some limitations to our analysis. 
In particular, we do not consider any selection effects, which for a population such as ours are two-fold.
First, intrinsically brighter short gamma-ray bursts are assumed to be observed on-axis and as such the emission produced by the interaction of the burst with the surrounding environment is brighter than the putative neutron star post-merger remnant. 
This implies that for on-axis short gamma-ray bursts, the window to infer the presence of a sharp drop due to the collapse of a long-lived neutron star is shorter as the initial emission from the jet has to drop to a level such that the emission from the neutron star can be observed.
Second, \swift{} typically takes up to~$\sim 100$ seconds to slew and observe an X-ray afterglow implying it will not see the collapse of some long-lived neutron stars that collapse before $\sim 100$ seconds. 
However, to complicate this further, this is the time measured in the detector frame which is red-shifted by an amount often not known. 
We aim to formulate and incorporate these selection effects in the future, however, we note that both these effects currently do not influence our results.
We have verified this with injection studies with up to $20$ events in our population and the bias caused by these effects is below our measurement uncertainty and will only become important as the population grows.

As described in Sec.~\ref{sec:methodology} we numerically marginalized over the unknown individual masses and spin periods of the putative post-merger remnants in our population. 
These marginalisations add uncertainty to our measurements as they propagate the uncertainty from not knowing these parameters into our inferred population parameters. 
In the future, with a possible coincident detection of gravitational waves from a binary neutron star inspiral and an X-ray afterglow, we can avoid these marginalization's or have a more informative prior, which will lead to a much more informative measurement.

In conclusion, we have shown that X-ray afterglow observations of short gamma-ray bursts can be used to constrain properties of post-merger remnants, with the population properties offering critical insight into the nuclear equation of state and gravitational-wave emission from newly born neutron stars.
In light of GW190425, we measure $\Mtov{} = 2.31 ^{+0.36}_{-0.21} M_{\odot}$ marginalised over all possible values of the mixing fraction describing the mass distribution of double and single neutron star systems in our galaxy, the latter being consistent with the progenitors of GW190425.
If instead, we assume a mixing fraction $\epsilon = 0$, i.e a mass distribution consistent with the double neutron star systems in our galaxy but one that rules out GW190425 having neutron star progenitors, we measure $\Mtov{} = 2.26 ^{+0.31}_{-0.17} M_{\odot}$.
\section{Acknowledgments}
We are grateful to Colm Talbot for helpful discussions on population inference.~We also thank Eric Thrane for his insightful comments on selection effects. This work made use of data supplied by the UK Swift Science Data Centre at the University of Leicester. N.S. is supported through an Australian Postgraduate Award. P.D.L. is supported through Australian Research Council Future Fellowship FT160100112 and ARC Discovery Project DP180103155.
\appendix*
\section{}\label{appendix}
\begin{figure}[ht]
\centering
\includegraphics[width=0.5\textwidth]{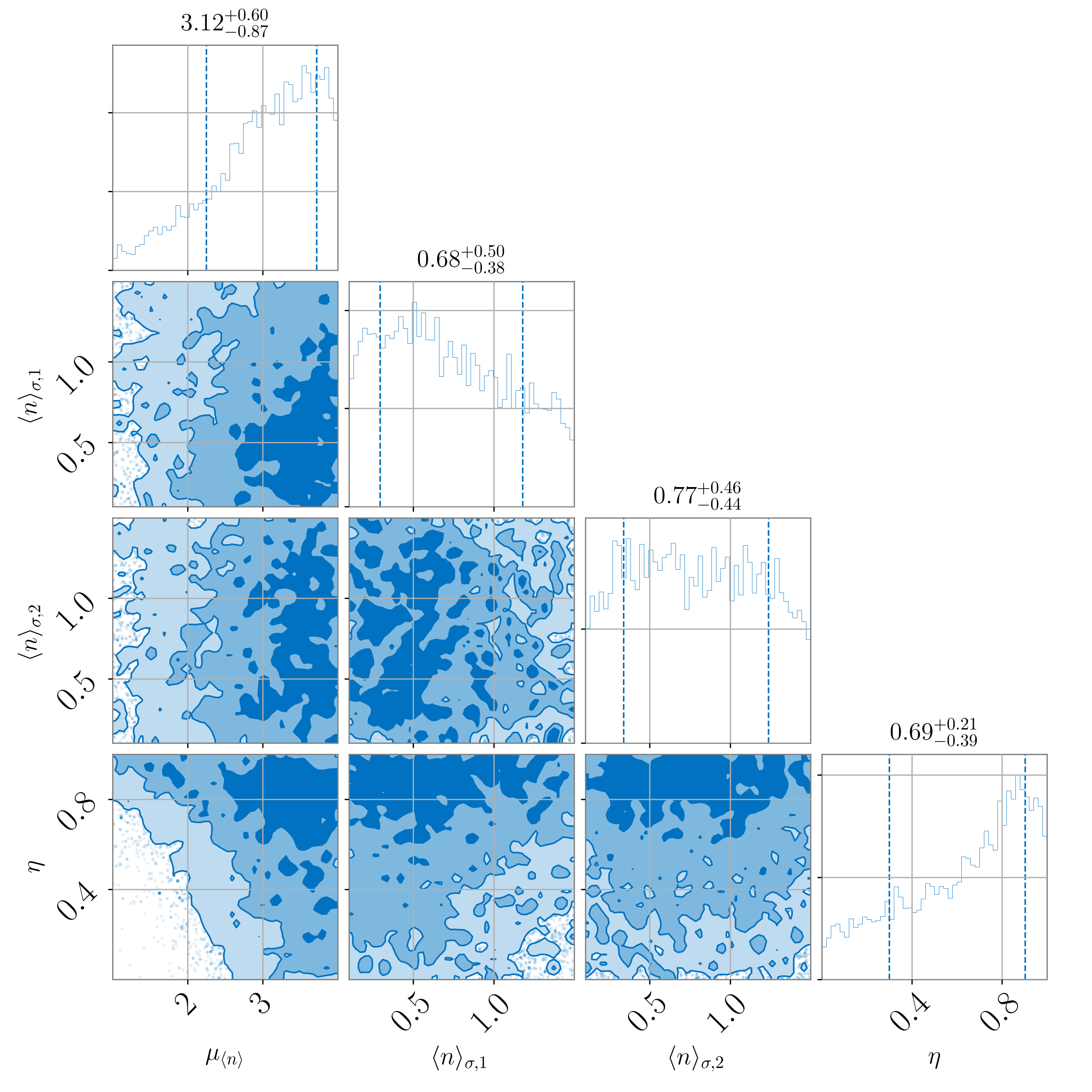}
  \caption{Corner plot showing the one and two-dimensional posterior distributions on $\mu_{\nn{}}$, $\nn{}_{\sigma,1}$, $\nn{}_{\sigma,2}$ and $\eta$. The shades of blue correspond to one-two-three sigma confidence intervals.}
  \label{fig:corner_brakingindex}
\end{figure}

\bibliographystyle{apsrev4-1} 
\bibliography{ref}
\end{document}